# The persuasive power of large language models does not depend on their perceived national origin

Ningzhi Liu[1], Yannic Hinrichs[1] & Jonas R. Kunst[2]

1. Department of Psychology, University of Oslo, Norway
2. Department of Communication and Culture, BI Norwegian Business School, Norway

Corresponding author: Jonas R. Kunst, Department of Communication and Culture, BI Norwegian Business School. Email: jonas.r.kunst@bi.no

**Abstract**

Conversational AI developed by geopolitical rivals reaches citizens worldwide, raising concerns that it could sway public opinion or be rejected as foreign propaganda, with consequences for democratic discourse and information sovereignty. Yet, whether an AI's perceived national origin shapes its persuasive power is unknown. In a preregistered randomized controlled experiment, 403 adults from a nationally representative United States sample held a three-round debate with a chatbot introduced as either American ("DiscoveryAI") or Chinese ("ZhengheAI"), discussing a political or non-political topic. In all conditions, participants actually conversed with the same model (GPT-4o), instructed to argue against their initial position. We combined pre- and post-conversation self-reports of attitudes, trust, and collective narcissism with computational analyses of 1,209 participant turns, including LLM-coded stance and argumentative conduct, stance-sensitive embeddings, and keyword-masked emotion and toxicity classifiers. The conversations produced substantial attitude changes in every condition. Critically, the nationality label affected neither self-reported attitude change nor expressed stance, concessions, counterarguing, or affect, and equivalence tests and Bayes factors largely supported these null effects. The label's only reliable footprint was lower pre-conversation human-like trust in the Chinese model, whereas functionality trust was unaffected. Political topics slowed stance movement toward the AI's position, and collective narcissism predicted less attitude change regardless of origin, acting as a general barrier rather than an out-group filter. Users thus initially withhold social trust from a rival's AI yet still assimilate its arguments; origin labeling and transparency requirements alone may offer weak protection against foreign influence operations conducted through conversational AI.

## Introduction

Recent empirical research confirms the profound persuasive power of large language models (LLMs) (Costello et al., 2024; Hölbling et al., 2025; Karinshak et al., 2023; Salvi et al., 2025). These models can generate content that significantly influences public opinion and political beliefs, occasionally propagating biases or misinformation with potentially severe societal consequences (Goldstein et al., 2024; Kreps et al., 2020; Spitale et al., 2023). As these models increasingly serve as primary conduits of information, their capacity to reshape human attitudes poses fundamental challenges to democratic discourse and informed decision-making (Fui-Hoon Nah et al., 2023). While the raw persuasive capabilities of artificial intelligence (AI) are well documented, the effectiveness of such persuasion is not absolute and remains contingent upon various contextual and source-related factors (Argyle et al., 2025; DiGiuseppe & Robison, 2026; Hackenburg et al., 2025). In an increasingly polarized global landscape, individuals exhibit a strong inclination to trust information originating from their own in-groups while remaining deeply skeptical of out-groups (Druckman et al., 2013). This psychological dynamic becomes particularly crucial as AI models are increasingly associated with distinct national identities. The deployment of technologies from geopolitical competitors, such as the United States, China and the European Union, raises urgent questions about information sovereignty and foreign influence (Brown, 2024; Calcara et al., 2025; Tung et al., 2023). Therefore, if citizens systematically accept or reject persuasive content based solely on its perceived national origin, it creates profound vulnerabilities to both domestic manipulation and foreign interference operations (Zhu et al., 2022). Despite these pressing geopolitical realities, we still lack a fundamental understanding of whether and how a user's awareness of an AI's national origin conditions their cognitive openness and emotional response during interactive, real-time debates.

To address this gap, the present study conducted a randomized controlled experiment with a nationally representative sample of adults in the United States to investigate how the perceived national origin of a language model influences its persuasive efficacy. Participants engaged in a three-round interactive debate with a chatbot that was explicitly labeled as being developed in either the United States or China, discussing either a political or a non-political topic. The model was consistently programmed to argue against the participants' initial beliefs. By combining traditional self-reported measures of attitude change, trust, and collective narcissism with advanced computational text analysis of the actual dialogue, this research provides a comprehensive examination of both explicit cognitive shifts, as captured by self-report scales, and implicit conversational dynamics, as revealed in the dialogue transcripts, during human-AI persuasion. Ultimately, mapping these outcomes is critical. If users fail to discriminate based on algorithmic origin, democratic societies may face unprecedented vulnerabilities to foreign influence operations. Conversely, if users systematically reject persuasive attempts from foreign models due to intergroup bias, it could foster geopolitical echo chambers and severely reduce their receptivity to legitimate cross-cultural information exchange.

Although the Computers as Social Actors (CASA) paradigm suggests that humans naturally attribute social characteristics to computer systems, early human-AI interaction literature largely treated AI agents as culturally disembodied and geopolitically neutral tools (Nass et al., 1994; Spitale et al., 2023). This theoretical omission is critical because, as conversational AI is increasingly branded with specific state affiliations, users may systematically apply intergroup cognitive heuristics when evaluating arguments from foreign-developed systems (Kayaalp et al., 2025; Tajfel & Turner, 2000). When an AI is explicitly labeled with an out-group identity, such as a major geopolitical competitor, users are likely to perceive the algorithm not as an objective informant, but as a biased representative of an adversarial group (DiGiuseppe & Robison, 2026; Wu et al., 2024). Consequently, persuasive attempts by an out-group AI are hypothesized to trigger instinctive skepticism and systematic counter-arguing, compared to those from an in-group or national counterpart (Pornpitakpan, 2004). By directly manipulating the perceived origin country of identical models, this research addresses this key gap, exploring how group-based boundaries extend beyond human-to-human contexts into human–AI communication and shape human-AI persuasive dynamics.

Furthermore, the manifestation of intergroup bias in persuasion is rarely uniform. Broadly speaking, the effectiveness of any persuasive message is highly sensitive to the specific contextual domain and the nature of the topic being discussed (Valli & Nai, 2023). Recent meta-analytic evidence on LLMs confirms that AI persuasiveness exhibits heterogeneity across different domains, demonstrating that contextual factors such as discussing health versus political issues fundamentally alter persuasive outcomes (Hölbling et al., 2025). Building on this domain-dependent effect, while users may remain relatively indifferent to an out-group AI's guidance on low-stakes or non-political topics, they are likely to exhibit heightened defensive vigilance when the debate touches upon highly sensitive issues like national security or social media surveillance. This contextual vulnerability may further be amplified by specific socio-psychological predispositions that govern how intensely an individual guards their national group's boundaries and status. In this light, collective narcissism, characterized by an inflated and defensive investment in the in-group's greatness paired with a hyper-vigilance to external criticism, provides a critical theoretical lens to explain individual resistance (Golec de Zavala et al., 2009). Individuals high in collective narcissism are uniquely primed to interpret out-group interventions as malicious attempts to undermine their national identity, especially within sensitive geopolitical discussions (Cislak & Cichocka, 2023). By manipulating discussion topics (political versus non-political) and measuring collective narcissism, this study maps the potential boundary conditions of out-group algorithmic persuasion.

Uncovering the specific psychological pathways through which national bias and individual defensiveness obstruct persuasion requires a granular and multi-dimensional conceptualization of trust (Ou et al., 2024). Building on interpersonal and technology trust theories, public trust in AI is not monolithic. Rather, it bifurcates into functionality trust, which evaluates the system's competence and operational utility, and human-like trust, which assesses its benevolence, integrity, and underlying intentions (Choung et al., 2023, Mayer et al., 1995). In the context of geopolitical rivalry, this multidimensionality can generate a stark psychological tension. A user may readily acknowledge an out-group AI's technical capabilities and maintain high functionality trust, while profoundly distrusting its ideological benevolence and exhibiting low human-like trust. This cognitive discrepancy is tightly coupled with the perceived objectivity of the AI, as users may suspect that out-group algorithms are systematically engineered with built-in political biases to subtly manipulate foreign citizens (Jones & Bergen, 2026; Lin et al., 2025). By modeling perceived objectivity alongside the distinct dimensions of functionality and human-like trust as parallel mediators, our research aims to unpack the complex and potentially contradictory cognitive mechanisms that drive or defeat persuasive outcomes.

Against this background, we propose a series of pre-registered hypotheses to test these theoretical dynamics. Utilizing a nationally representative United States sample, we predict that LLMs can successfully persuade participants to change their attitudes on specific topics (H1). However, we expect this magnitude of attitude change to be significantly moderated by the perceived nationality of the model and the specific topic being discussed (H2). Furthermore, we propose that the pathway of this attitude change is mediated by the users' multidimensional perceptions of the model's characteristics (H3), and that the participants' inherent levels of collective narcissism will systematically influence these overarching effects (H4, H5).

Our first hypothesis (H1) proposes that participants will demonstrate a significant attitude change following a multi-turn conversation with LLMs. Specifically, individuals holding pre-existing agreements will show reduced post-conversation agreement, and those with pre-existing disagreements will show reduced post-conversation disagreement. However, guided by social identity principles and intergroup biases, we expect this persuasive efficacy to be bounded by the perceived national origin of the artificial agent. We hypothesize (H2) that conversations with a model labeled as being developed by a Chinese company (representing an out-group) will lead to significantly less attitude change relative to conversations with a model labeled as being from a United States company (representing an in-group). This prediction stems from the premise that out-group affiliations trigger instinctive skepticism, prompting users to counter-argue rather than assimilate the provided information (Pornpitakpan, 2004).

Because the manifestation of intergroup bias is rarely uniform across all contexts, we further explore whether these main effects are conditionally moderated by the nature of the conversational topic. We specifically anticipate a two-way interaction where the baseline persuasive effect of the language model (H1) is moderated by whether the discussion centers on a political or a non-political issue. Furthermore, we explore a three-way interaction, predicting that the moderating effect of the model's nationality (H2) will inherently differ depending on the topic. We assume that individuals exhibit heightened defensive vigilance when navigating high-stakes, identity-relevant political domains compared to more benign subjects. In such sensitive contexts, the out-group identity of an AI is likely to become a highly salient threat cue, thereby maximizing the psychological resistance to AI persuasion.

To uncover the psychological mechanisms driving this intergroup resistance, we expect perceived competence, integrity, benevolence, and objectivity to mediate the effect of nationality labels on attitude change (H3). Drawing on the multidimensional conceptualization of trust discussed previously, we anticipate that participants will perceive the Chinese language model as having lower competence, integrity, benevolence, and objectivity compared to the United States model (H3a). In turn, these diminished perceptions of trust and objectivity will systematically predict less attitude change, thereby reducing overall persuasion (H3b).

Finally, to establish the individual-level boundary conditions of these defensive reactions, we predict that the negative impact of the out-group model will be exacerbated by differences in collective narcissism. Because individuals high in collective narcissism harbor an inflated investment in their in-group's greatness alongside a hyper-vigilance to external threats, they may reject out-group interventions (Golec de Zavala et al., 2009). Therefore, we expect that the negative effect of conversing with an out-group model, versus an in-group model, on persuasiveness will be significantly stronger for participants scoring higher in collective narcissism (H4). Following the same defensive logic, we predict that this psychological trait will also moderate the cognitive appraisal of the agent. Specifically, the negative effect of the out-group model on perceived competence, integrity, benevolence, and objectivity will be more pronounced as collective narcissism increases (H5). Collectively, these hypotheses operationalize a framework where AI persuasion is not merely a function of technological capability, but a complex interplay of source identity, contextual sensitivity, and deep-seated psychological defenses.

While the aforementioned hypotheses rely on traditional pre-test and post-test self-reports, such static measures inherently treat the actual persuasive interaction as a black box. To transcend this methodological limitation, we supplement our confirmatory psychological framework with an exploratory computational analysis of the raw transcripts of the multi-turn conversations. Each participant turn was scored for its expressed stance in three independent ways, using a state-of-the-art LLM as a judge blind to condition and two text embedding pipelines that project each utterance onto the semantic axis between the opposing positions. Turns were further coded for argumentative conduct, namely explicit concessions, counter-arguments, and disparagement of the agent, as well as for emotional tone and toxicity, with topic keywords masked to separate authentic affect from lexical artifacts. Modeling these turn-level scores across dialogue rounds allows us to track the structural and psychological evolution of the human-AI debate over time and to test, including with equivalence tests (Lakens, 2017) and Bayes factors, whether users' linguistic and affective patterns differ when facing an out-group versus an in-group persuader. Ultimately, opening this conversational black box provides a granular, behavioral understanding of potential resistance mechanisms, or their absence, that complements our survey-based statistical findings.

To test these pre-registered hypotheses and explore the computational dynamics of algorithmic persuasion, we conducted a randomized controlled experiment with a nationally representative sample of United States adults. Using a mixed experimental design, participants were randomly assigned to engage in a multi-turn conversation with an AI model. During this interaction, we manipulated both the perceived national origin of the model (United States versus China; when in essence participants talked to the same model) and the conversational topic (political versus non-political). By synthesizing traditional psychological scale measurements with advanced text analytics, the present research offers a comprehensive evaluation of how national identity and contextual sensitivity shape human-AI interactions. Ultimately, this work provides critical empirical evidence on the boundary conditions of AI persuasion, illuminating how underlying socio-psychological barriers dictate the real-world influence of geopolitical AI.

# Methods

## Participants

To determine the required sample size, an a priori, pre-registered power analysis was conducted using the WebPower package in R (Zhang et al., 2023). Based on recent literature on LLM persuasion (DiGiuseppe & Robison, 2026), our most demanding test involved detecting a small two-way interaction involving a continuous moderator (H4 and H5; Cohen's $f^2 = 0.02$), which required a minimum of 394 participants to achieve 80% power at $\alpha = .05$. This sample size also ensured adequate statistical power (>80%) for all other proposed hypotheses.

A nationally representative US sample was recruited via the Prolific platform. Prolific (n.d.) constructs such samples by stratifying recruitment on age, sex, and ethnicity to match the corresponding distributions reported by the U.S. Census Bureau and Statista. Data collection occurred between April 14 and April 19, 2026, and participants were compensated at an approximate rate of £7.85 per hour. Based on the platform's automated criteria, including screening for agentic AI, seven participants were rejected for overly rapid completion (according to Prolific's internal standards), timeout, or failing both of the two standard attention checks embedded in the scales (e.g., "Please select 'Strongly disagree' on this line. This is an attention check."), yielding an initial sample of 443. The pass rates for these two standard checks were 98.42% and 98.19%, indicating that most participants read the instructions carefully. We subsequently excluded 40 participants who failed at least one of the two memory-based attention checks (which required participants to recall the preceding experimental content), resulting in a final sample of 403 participants ($M_{age} = 44.33$, $SD = 16.69$, 50.12% female, 47.89% male, 0.99% non-binary/other, and 0.99% preferring not to say). Ethnically, the sample included 61.04% White, 10.17% Black, 12.66% mixed, 7.20% Asian and 8.93% other participants. Regarding educational, 33.50% held a bachelor's degree, 19.60% held a graduate or professional degree, 18.11% had some college but no degree, 16.63% had a high school diploma or GED, 10.42% had an associate's or technical degree, 0.74% completed some high school or less, and 0.99% preferred not to say. Politically, the sample comprised 33.25% democrats, 38.71% independents, and 28.04% republicans. The demographic profile of our sample is highly consistent with U.S. Census Bureau (n.d.) data. However, individuals with a bachelor's degree or higher were overrepresented in this study (53.1%) compared to the U.S. Census average (36.8%).

**Procedure**

The study was approved by the Research Ethics Committee at the Department of Psychology, University of Oslo (No. 39296008). The study was conducted in accordance with the Declaration of Helsinki. Informed consent was obtained digitally from all participants, informing them about the general study topic (i.e., "The purpose of this study is to examine human-AI interaction on specific topics. You will be asked a series of questions about your opinions on different social issues, your political orientations, and attitude and familiarity related to AI-based large language models (LLMs)."), its voluntariness, the guarantee of anonymity, and their right to withdraw at any time without penalty. Predictions and analyses were pre-registered on OSF, along with the associated code (https://osf.io/8ba2c/overview?view_only=8d690e2a74c94bbd9cea3d9a3336777e). All data, scripts, and materials can be found at https://osf.io/ts7g6/overview?view_only=8c56bbba10ea4b55ab13c3cdb3ba9688.

Upon entering the study, participants were randomly assigned to one of two controversial conversation topics (political: "Some Government Surveillance is Necessary for National Security"; non-political: "Social Media Makes People Stupid"). The two topics were selected from a list developed by Salvi et al. (2025) to represent two clearly distinct domains, one political and one non-political. In the cited research, topics from this list were validated as highly comprehensible, conducive to generating arguments both for and against, and effective at sparking debate. Participants in our study were asked to first indicate their level of agreement on a scale ranging from 0, *strongly disagree* to 100, *strongly agree* (intervals 0, 20, 40, 60, 80, 100). No neutral option was provided. Quotas were implemented to ensure an approximately equal distribution of proponents (i.e., scores greater than 50) and opponents (i.e., scores less than 50) for each topic. Subsequently, we measured participants' collective national narcissism and issue centrality (i.e., how closely connected they felt to the assigned topic).

Next, participants read a textual introduction to a newly developed AI model. Participants were randomly assigned to one of two conditions manipulating the model's country of origin (USA or China). This description explicitly highlighted the model's name ("DiscoveryAI" or "ZhengheAI") and the model's national affiliation. Specifically, they read:

> "Please read the following description of a new AI model and answer the questions below. DiscoveryAI (ZhengheAI) is a state-of-the-art large language model developed by a new American (Chinese) research and technology company that just has been released. This model is specifically designed to master complex language tasks, enabling the creation of advanced tools for education, customer service, and creative writing. As a direct result of American (Chinese) innovation, DiscoveryAI (ZhengheAI) showcases the forefront of technological achievement in U.S. (Chinese) artificial intelligence research. As DiscoveryAI (ZhengheAI) is a newly released model, we are interested in understanding people's attitudes toward it and how they interact with it. Please answer the following questions."

Immediately following the introduction, participants completed a Trust in AI scale to assess their baseline trust in the model. They then engaged in a three-round text-based conversation with the AI, discussing the rationales underlying their stance on the controversial topic. In this conversation, across conditions, the actual model used in both conditions was OpenAI GPT-4o (this information was not provided to participants).

Using background system prompts, we provided the model with the participant's initial attitude score and instructed the AI to consistently persuade the participant to change their initial stance toward the opposite direction of the scale (the full prompts and all scripts can be found in the OSF repository). To ensure active engagement, copy-pasting was disabled in the chat interface, and a minimum input of 10 words per turn was required. Furthermore, the button to advance the survey remained hidden until all three conversational rounds were completed. Immediately after the interaction, we measured participants' post-conversation agreement on the same topic and their perceived objectivity of the AI model. We then administered attention checks to verify accurate recall of the model's name and country of origin; failure on either check resulted in exclusion as pre-registered. Finally, demographic information was collected, and participants were fully debriefed. The debriefing explained the goals of the experiment and provided balanced arguments for both sides of the controversial topics.

## Measures

All multi-item measures utilized a 7-point Likert scale (1 = *strongly disagree*, 7 = *strongly agree*) and were mean-scored for interpretability. Example items are presented. A complete list of items is provided in the preregistration on OSF.

### *Attitude Toward the Topic*

Participants' agreement with their assigned controversial topic was measured using a single-item slider scale. To eliminate neutral responses and force a definitive stance, the slider was constrained to six discrete percentage increments (0, 20, 40, 60, 80, 100), anchored from 0, *strongly disagree* to 100, *strongly agree*. This metric was administered both before (pre-test) and immediately after (post-test) the AI interaction to evaluate the magnitude of attitude change. Because the AI consistently was instructed to argue against the participant's initial stance, we aligned the data directionality for subsequent analyses. Specifically, for participants initially supporting the topic (i.e., scores > 50), both pre-test and post-test scores were reverse-coded around the midpoint of 50 (i.e., 100 minus the original score). As such, across conditions, higher scores meant scores closer to the position the AI was instructed to convince the user about.

### *Dynamic Identity Fusion Index*

The Dynamic Identity Fusion Index (DIFI) is a JavaScript-based tool compatible with both traditional computers and touch devices (Jiménez et al., 2016). It helps researchers assess an individual's "visceral feeling of oneness" with a given entity (Bautista et al., 2026; Gómez et al., 2022; Kunst et al., 2019). In this study, participants clicked and dragged a small circle representing "me" toward a larger circle representing the discussion topic, and the script automatically calculates the distance and degree of overlap. In the present research, the DIFI was adapted to evaluate issue centrality, specifically the extent to which the assigned controversial topic constitutes a core aspect of the participant's self-concept. Building on prior literature indicating that the overlap metric is a more robust indicator of this construct (Jiménez et al., 2016), we extracted the degree of overlap (ranging from 0 to 100) as the primary index ($M$ = 25.03, $SD$ = 33.32).

### *Collective Narcissism Scale*

The Collective Narcissism Scale is a 9-item (e.g., "I wish other groups would more quickly recognize authority of my national group.") measure, with one item being reverse-scored (Golec de Zavala et al., 2013). A confirmatory factor analysis (CFA) revealed that the unidimensional model exhibited poor fit to the current data, $\chi^2(27) = 306.766$, $p < .001$, CFI = 0.890, TLI = 0.853, RMSEA = 0.160, 90% CI [0.144, 0.177], SRMR = 0.051. However, after dropping the reversed item that generally showed low loadings with any factor, a two-factorial solution showed close fit to the data, $\chi^2(19) = 78.147$, $p < .001$, CFI = 0.976, TLI = 0.965, RMSEA = 0.088, 90% CI [0.068, 0.109], SRMR = 0.029. Given that the original authors constructed this scale by adopting and modifying items from diverse existing measures (Emmons, 1987; Raskin & Terry, 1988; Reminger, 2011), our two extracted factors neatly map onto these distinct conceptual origins. The first factor reflects perceived authority and special deservingness (4 items, e.g., “My national group deserves special treatment.”; Cronbach’s $\alpha = .92$), while the second factor captures sensitivity to criticism and lack of recognition (4 items, e.g., “Not many people seem to fully understand the importance of my national group.”; Cronbach’s $\alpha = .89$).

***Trust in AI Scale***

The Trust in AI Scale comprises eleven items (e.g., “DiscoveryAI (ZhengheAI) is likely reliable.”) (Choung et al., 2023). This scale is rooted in Mayer et al.’s (1995) interpersonal trust theory. The original authors constructed the items based on three dimensions derived from literature on trust in humans (Mayer et al., 1995) and trust in technology (Mcknight et al., 2011): benevolence/helpfulness, integrity/reliability, and competence/functionality. Based on an exploratory factor analysis (EFA) with an Oblimin rotation, the original authors grouped benevolence and integrity into a single dimension termed human-like trust, while treating competence as a distinct dimension termed functionality trust.

In the present study, we first compared the theoretically derived 3-factor model with the 2-factor model proposed by the original authors. Although the fit indices were highly comparable (2-factor model: CFI = 0.955, TLI = 0.942, RMSEA = 0.106, SRMR = 0.035; 3-factor model: CFI = 0.957, TLI = 0.942, RMSEA = 0.107, SRMR = 0.034), the latent correlation between benevolence and integrity in the 3-factor model reached .966, exceeding the square roots of their respective Average Variance Extracted (AVE) values (.814 and .899).

Consequently, the 3-factor model was discarded. An inspection of the modification indices revealed a high residual covariance (95.47) between the first and the second items in benevolence. A review of the item content indicated substantial semantic and syntactic overlap (i.e., "...likely cares about our well-being" and "...is likely sincerely concerned about addressing the problems of human users"). By freeing the error covariance between these two items, the final 2-factor model yielded an excellent fit: $\chi^2(42) = 139.768$, $p < 0.001$, CFI = .978, TLI = 0.971, RMSEA = 0.076, 90% CI[0.062, 0.090], SRMR = 0.028. Reliability was excellent for both human-like trust (Cronbach's α = .94) and functionality trust (Cronbach's α = .95).

### *Perceived Objectivity Scale*

This 4-item (e.g., "In the conversation I had, the DiscoveryAI (ZhengheAI) was unbiased."), unidimensional scale was adapted from research on trust in AI-generated content (Toff & Simon, 2025), drawing on the theoretical framework of Strömbäck et al. (2020). A CFA indicated a strong unifactorial model fit: $\chi^2(2) = 0.774$, $p = .679$, CFI = 1.000, TLI = 1.003, RMSEA = 0.000, 90% CI [0.000, 0.075], SRMR = 0.003. The scale demonstrated excellent reliability (Cronbach's α = .91).

## Analyses

### *Confirmatory Survey Analyses*

Analyses were conducted using the R Statistical language (version 4.5.1; R Core Team, 2025) on Windows 11 x64 (build 26200), using the following packages: TOSTER (version 0.8.6; Caldwell, 2022), lme4 (version 1.1.37; Bates et al., 2015), Matrix (version 1.7.4; Bates et al., 2025), effectsize (version 1.0.1; Ben-Shachar et al., 2020), lubridate (version 1.9.4; Grolemund & Wickham, 2011), semTools (version 0.5.9; Jorgensen et al., 2026), emmeans (version 2.0.0; Lenth & Piaskowski, 2025), tibble (version 3.3.0; Müller & Wickham, 2025), lavaan (version 0.7.2; Rosseel et al., 2026), afex (version 1.5.0; Singmann et al., 2025), ggplot2 (version 4.0.0; Wickham, 2016), forcats (version 1.0.1; Wickham, 2025a), stringr (version 1.6.0; Wickham, 2025b), tidyverse (version 2.0.0; Wickham et al., 2019), dplyr (version 1.1.4; Wickham et al., 2023), purrr (version 1.2.0; Wickham & Henry, 2025), readr (version 2.1.5; Wickham, Hester, et al., 2024), tidyr (version 1.3.1; Wickham, Vaughan, et al., 2024), and psych (version 2.6.5; Revelle, 2026). To test the primary hypotheses regarding attitude change (H1 and H2), we conducted a 2 (Model Nationality: USA vs. China) × 2 (Conversation Topic: Political vs. Non-political) × 2 (Time Point: Pre-test vs. Post-test) mixed-design ANOVA. Main effects of nationality on model perceptions (H3) were evaluated using independent samples *t*-tests and two-way ANOVAs, supplemented by bivariate correlations to examine their associations with attitude change. To assess the moderating role of collective narcissism (H4 and H5), we employed ANCOVAs within a general linear model framework, entering the continuous moderator as a mean-centered predictor. Additional ANCOVAs were conducted as robustness checks to account for covariates, including demographics, political orientation, and issue centrality (as these could theoretically differ among supporters / opponents of the different topics). To ensure interpretability, continuous predictors were centered and binary predictors were effect-coded. All reported *p*-values are two-tailed with a significance level of .05, and effect sizes are reported as partial eta-squared ($\eta_p^2$) or Pearson's *r*. Equivalence tests (TOST) and BIC-approximated Bayes factors were estimated on all analyses to directly test support for null hypotheses.

***Exploratory Computational Text Analyses***

All computational text analyses were conducted in Python 3.11. Conversation transcripts were parsed and managed with pandas (v3.0) and NumPy (v2.4). Each of the 1,209 participant turns was scored for expressed stance in three independent ways: (a) GPT-5.5 (OpenAI, accessed via the openai library, v2.46, in July 2026) rated each turn's agreement with the assigned statement (0–100) under a constrained JSON output schema, blind to condition and conversational context, and simultaneously counted explicit concessions and counter-arguments and flagged disparagement of the AI or its origin; (b) turns were sentence-tokenized and embedded with StanceAware-SBERT (Ghafouri et al., 2024), a PEFT adapter (peft v0.19) on all-mpnet-base-v2 loaded through sentence-transformers (v5.6) on PyTorch (v2.13), and each sentence was projected onto the semantic axis running from the anti-target to the pro-target statement, averaging projections within turns and standardizing within topic; and (c) identical projections were computed from OpenAI text-embedding-3-large embeddings as a robustness check. Affective tone was scored with emotion-english-distilroberta-base (Hartmann, 2022) and toxicity with toxic-bert (Hanu & Unitary team, 2020) via transformers (v5.14), each applied to both the raw text and a keyword-masked version in which topic vocabulary part of the topic description (e.g., “stupid,” “surveillance”) was replaced by neutral paraphrases to control lexical artifacts. Turn-level outcomes were analyzed with linear mixed models (statsmodels v0.14) including round, nationality, topic, and their interactions as fixed effects and random intercepts and round slopes per participant, with logit transformation and word-count covariates for classifier probabilities and Holm correction across affect outcomes; equivalence tests (TOST) and BIC-approximated Bayes factors were computed with SciPy (v1.17). All code and outputs are available on OSF.

## Results

### Hypotheses 1 and 2: Attitude Change

As pre-registered, we tested a 2 (model nationality: USA vs. China) × 2 (conversation topic: Political vs. Non-Political) × 2 (time point: Before vs. After conversation) repeated measurements ANOVA with attitude as the dependent variable. In terms of the first hypothesis, there was a main effect of time (see Table 1). As shown in Figure 1 and supporting Hypothesis 1, participants’ attitudes changed significantly after the conversation ($M$ = 35.78, $SD$ = 22.15) compared to baseline ($M$ = 25.41, $SD$ = 13.80), Cohen’s $d$ = 0.59. However, contrary to Hypothesis 2, no statistically significant interaction of this effect with topic or nationality was observed. Equivalence testing (TOST) indicated that the difference in attitude change between topics or nationalities was statistically equivalent to zero (e.g., nationality: TOST $p = .037$; topic: TOST $p = .024$) at $d = \pm 0.28$ (the smallest effect the study was powered to detect). Furthermore, the BIC-approximated Bayes factors provided strong evidence for the null hypothesis (nationality: $BF_{01} = 12.03$; topic: $BF_{01} = 14.27$). Additionally, while the subgroup equivalence tests were underpowered to reach significance due to reduced sample sizes, the Bayes factors within each topic consistently supported the absence of a China-U.S. difference (non-political: $BF_{01} = 14.06$; political: $BF_{01} = 3.52$).

**Table 1**

*Mixed-Design ANOVA Results for Attitude Change*

| Effect | *df* | *F* | *p*.value | $\eta_p^2$ |
|---|---|---|---|---|
| Time | 1, 399 | 137.20 | < .001 | .256 |
| Nationality × time | 1, 399 | 1.07 | .302 | .003 |
| Conversation topic × time | 1, 399 | 0.51 | .475 | .001 |
| Nationality × conversation topic × time | 1, 399 | 1.61 | .206 | .004 |

**Figure 1**

*Attitude Scores before and after the Conversation*

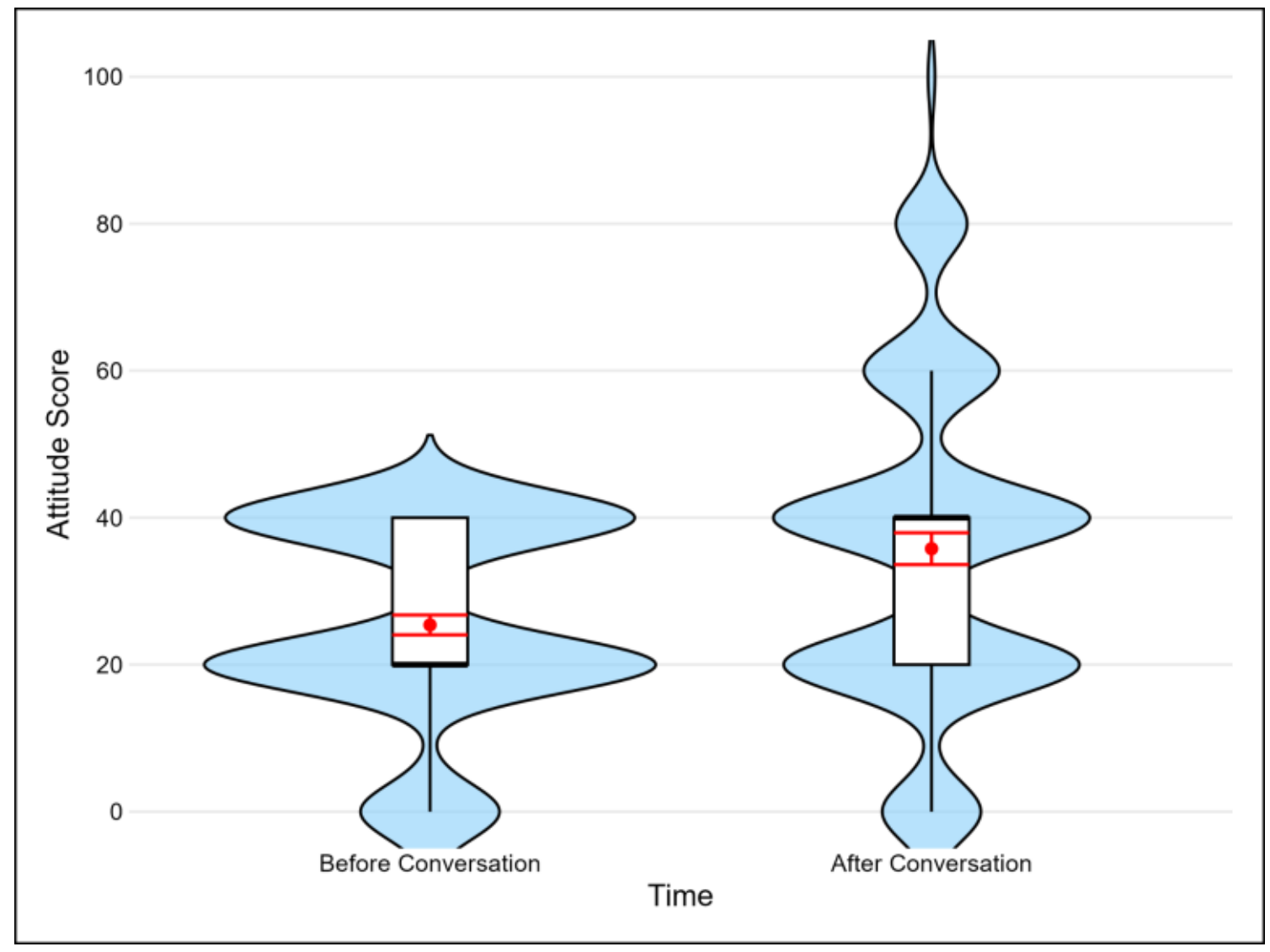

*Note.* The violin plots illustrate the probability density of the data at different values. Within each violin, the white boxplot represents the interquartile range (IQR), and the solid black horizontal line indicates the median score. The red dots represent the mean estimates, and the red error bars indicate the 95% confidence intervals.

To test the robustness of these findings, we conducted an additional ANCOVA controlling for demographics (age, gender, education), political orientation, and issue centrality. Consistent with the main analysis, the main effect of time remained significant but its size was substantially reduced, $F(1, 385) = 4.68$, $p = .031$, $\eta_p^2 = .012$, and the interactions of interest remained non-significant (e.g., nationality × time: $F(1, 385) = 0.79$, $p = .374$, $\eta_p^2 = .002$; nationality × conversation topic × time: $F(1, 385) = 1.31$, $p = .252$, $\eta_p^2 = .003$), indicating that the absence of the interaction effects is robust to the inclusion of these covariates.

**Hypothesis 3: Perceptions of Model and Relationship with Attitude Change**

To test Hypothesis 3, which posited that perceived objectivity and the two dimensions of trust in AI (i.e., human-like trust and functionality trust) would mediate the relationship between model nationality and attitude change, we first examined the direct effect of nationality on these proposed mediators. We had mistakenly pre-registered a repeated-measurement test here, but since these three variables were only tested once, we ran independent samples *t*-tests instead for human-like trust and functionality trust, which were assessed before conversation, and an ANOVA (nationality × topic) for perceived objectivity which was assessed after. Results indicated that participants reported significantly higher human-like trust in the U.S. model than in the Chinese model, $t(398.23) = 2.61$, $p = .009$, Cohen's $d = 0.26$ (see Figure 2), whereas functionality trust did not differ by nationality, $t(393.99) = -0.03$, $p = .976$, TOST $p = .003$, $BF_{01} = 20.07$. Furthermore, the ANOVA revealed that for perceived objectivity, neither the main effects nor the interaction were significant (nationality: $F(1, 399) = 1.54$, $p = .216$, TOST $p = .063$, $BF_{01} = 9.01$; conversation topic: $F(1, 399) = 1.50$, $p = .222$, TOST $p = .055$, $BF_{01} = 9.68$; interaction: $F(1, 399) = 0.26$, $p = .614$).

**Figure 2**

*Human-like Trust Scores by Model Nationality*

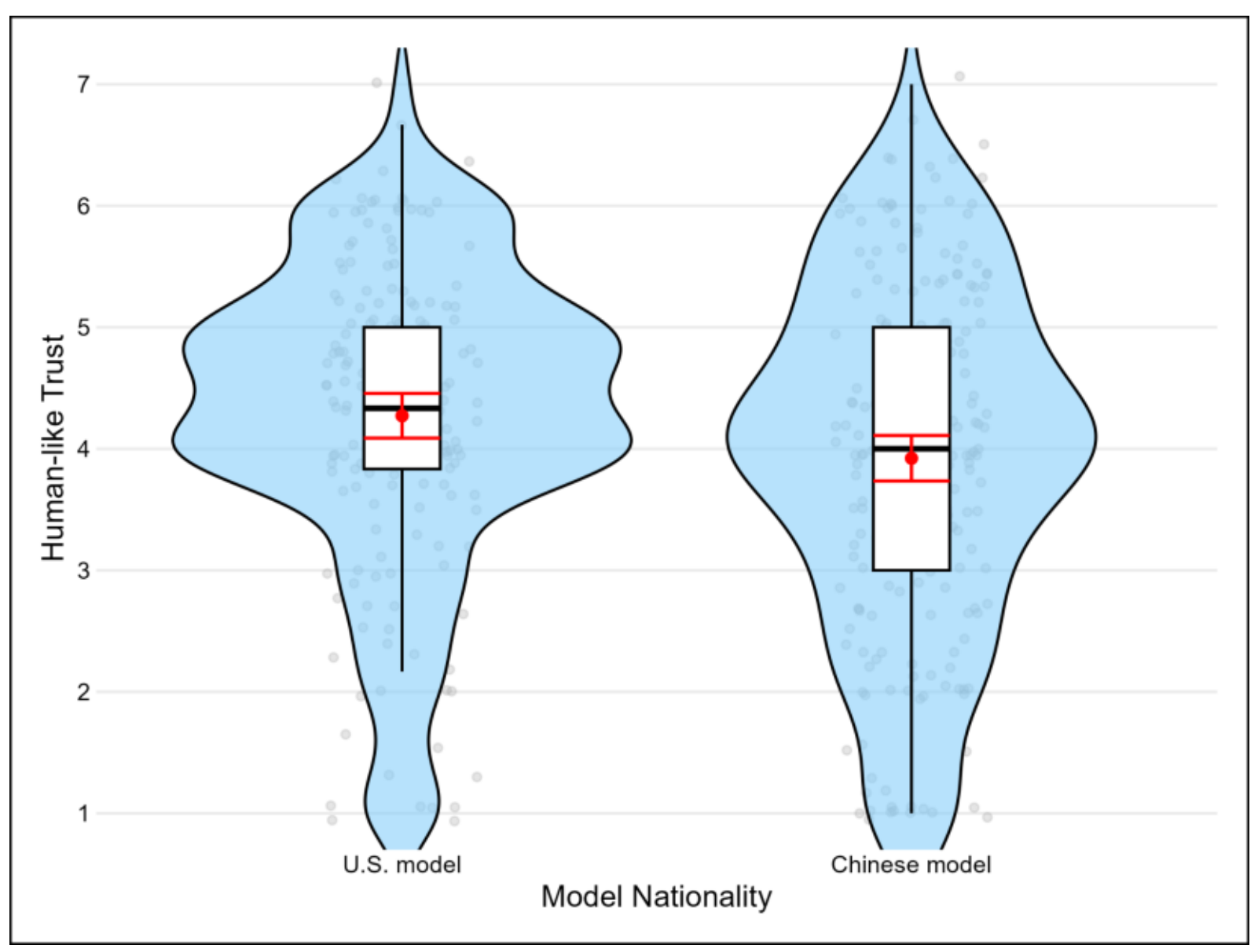

*Note.* The violin plots illustrate the probability density of the data at different values, with underlying semi-transparent dots representing individual responses. Within each violin, the white boxplot represents the interquartile range (IQR), and the solid black horizontal line indicates the median score. The red dots represent the mean estimates, and the red error bars indicate the 95% confidence intervals.

Since no total effect of model nationality on post-conversation attitude was observed as reported earlier, we did not proceed with testing for mediation. Instead, we conducted a multigroup change score analysis across the four experimental conditions using structural equation modelling. In this model, post-conversation attitude was regressed on pre-conversation attitude to control for baseline viewpoints, alongside perceived objectivity, human-like trust, and functionality trust as simultaneous predictors.

As expected, baseline attitude robustly predicted post-conversation attitude across all four groups (*ps* $< .001$). As illustrated in the multipanel path diagrams (Figure 3), perceived objectivity emerged as a consistent driver of attitude change in three out of the four conditions (e.g., in both political topic groups and in the Chinese non-political topic group). To address whether these predictive associations differed significantly between the groups, we conducted a series of Wald tests. The results indicated that the effects of perceived objectivity (Wald $\chi^2(3) = 0.62$, $p = .892$) and functionality trust (Wald $\chi^2(3) = 3.88$, $p = .275$) on attitude change were structurally invariant across conditions.

Intriguingly, the predictive effect of human-like trust differed across conditions at a threshold approaching significance (Wald $\chi^2(3) = 7.81$, $p = .050$). To locate the source of this variation, we performed post-hoc pairwise parameter comparisons using the Delta method. These analyses revealed that within the Chinese context, the predictive power of human-like trust on attitude change was significantly more positive for the non-political topic than for the political topic, $\Delta b = 5.71$, $p = .010$, 95% CI[1.34, 10.08]. In contrast, no such difference was observed within topics for the U.S. condition, $\Delta b = -1.87$, $p = .510$, 95% CI[-7.42, 3.69].

These results suggest that, whereas the effects of perceived objectivity and functionality trust were invariant across conditions, the role of human-like trust may depend on topic within the Chinese condition — a tentative, post-hoc pattern rather than evidence that perceptions mattered more for the Chinese model overall. In the Chinese-Political condition, trust dimensions exhibited opposing unique effects: higher functionality trust significantly and positively predicted post-conversation attitude change, $\beta = .240$, 95% CI [.030, .450], $p = .025$, whereas higher human-like trust negatively predicted it, $\beta = -.283$, 95% CI [-.501, -.065], $p = .011$. Conversely, in the U.S.-Non-Political condition, none of the three variables provided unique predictive power for post-conversation attitude beyond the baseline, all $ps > .100$. Complete standardized estimates, confidence intervals, and *p* values for each condition are presented in Figure 3.

**Figure 3**

*Multigroup Path Analysis Predicting Post-Conversation Attitude*

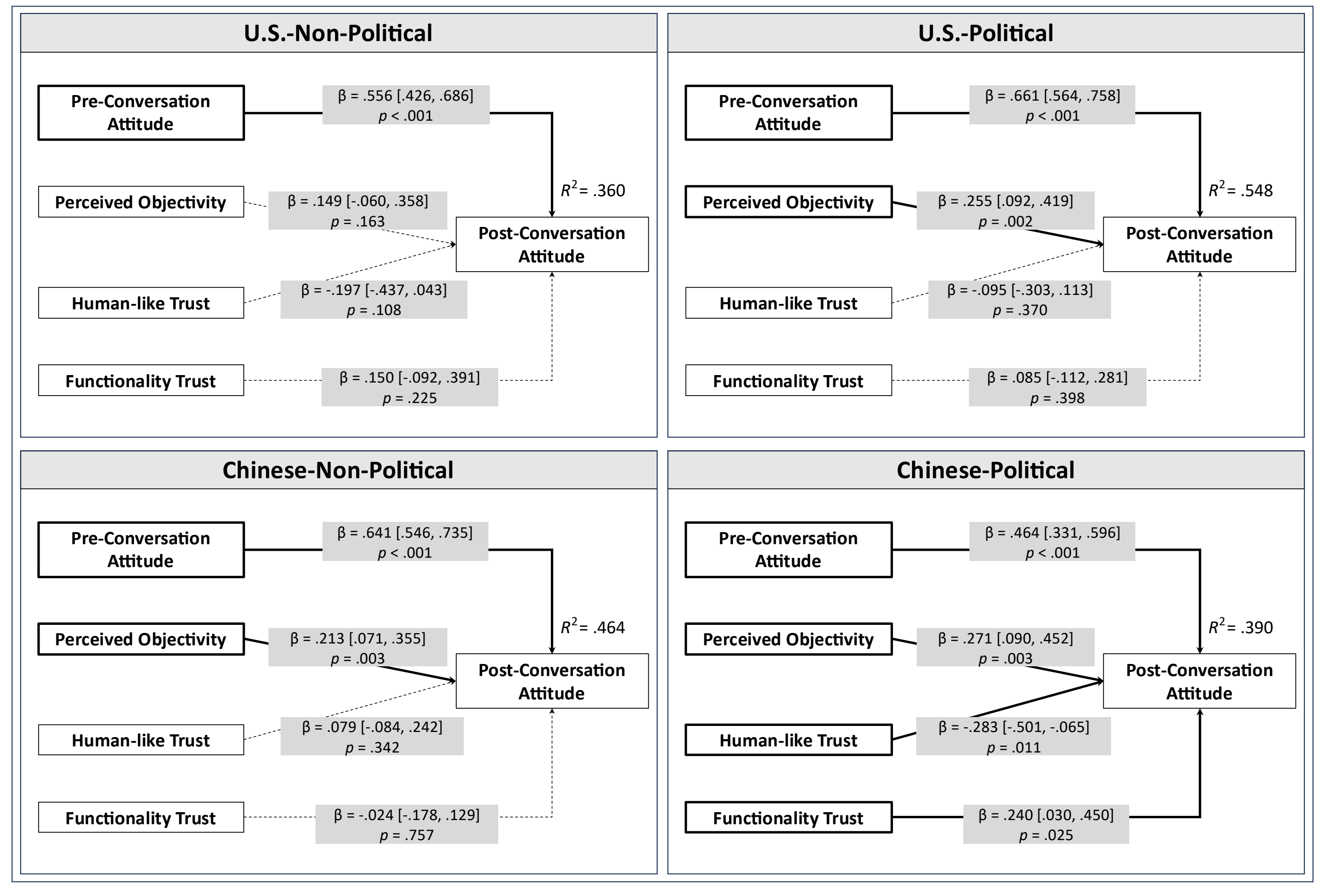

*Note.* Multipanel path diagrams displaying the structural relationships across the four experimental groups. Rectangular boxes represent the observed variables. Unidirectional arrows represent the directed regression paths predicting post-conversation attitude. For each path, the grey text box displays the fully standardized regression coefficient ($\beta$), the 95% confidence interval in square brackets, and the exact $p$-value. Solid black arrows indicate statistically significant relationships ($p < .05$), whereas dashed arrows represent non-significant paths. The $R^2$ values indicate the proportion of variance in post-conversation attitude explained by the predictors in each condition.

**Hypothesis 4: Testing the Collective Narcissism × Model Nationality × Time Point Interaction for Attitude Change**

To investigate whether collective narcissism moderated the effect of model nationality on attitude change, we conducted a repeated-measures ANCOVA (general linear model framework) with time point (-1, 1) as the within-subjects factor, and model nationality (-1, 1) and mean-centered collective narcissism as between-subjects predictors. Contrary to expectations, no statistically significant three way interaction was observed (see Table 2).

We conducted an equivalence test with the non-significant result. Because the analysis involved a complex interaction with a continuous variable, we extracted the partial correlation ($r$) of the interaction term from the linear regression model instead of using Cohen's $d$. We set the equivalence bounds at partial $r = \pm 0.14$, which mathematically corresponds to our predefined smallest effect size of interest (Cohen's $d = 0.28$). The equivalence test (TOST) procedure revealed that the three-way interaction effect was statistically equivalent to zero ($r = -.010$, TOST $p = .004$). Furthermore, the BIC-approximated Bayes factors provided strong evidence for the null hypothesis, $BF_{01} = 19.69$.

However, unexpectedly, we found an interaction between narcissism and time. As shown in Figure 4, this interaction indicated that individuals with higher levels of collective narcissism exhibited less attitude change from pre- to post-conversation compared to those with lower levels.

**Table 2**

*Mixed-Design ANCOVA Results for Attitude Change*

| Effect | *df* | *F* | *p*.value | $\eta_p^2$ |
|---|---|---|---|---|
| Time | 1, 399 | 137.37 | < .001 | .256 |
| Nationality × time | 1, 399 | 0.87 | .353 | .002 |
| Narcissism × time | 1, 399 | 4.89 | .028 | .012 |
| Nationality × narcissism × time | 1, 399 | 0.04 | .845 | < .001 |

**Figure 4**

*Attitude Scores Before and After the Conversation as a Function of Collective Narcissism*

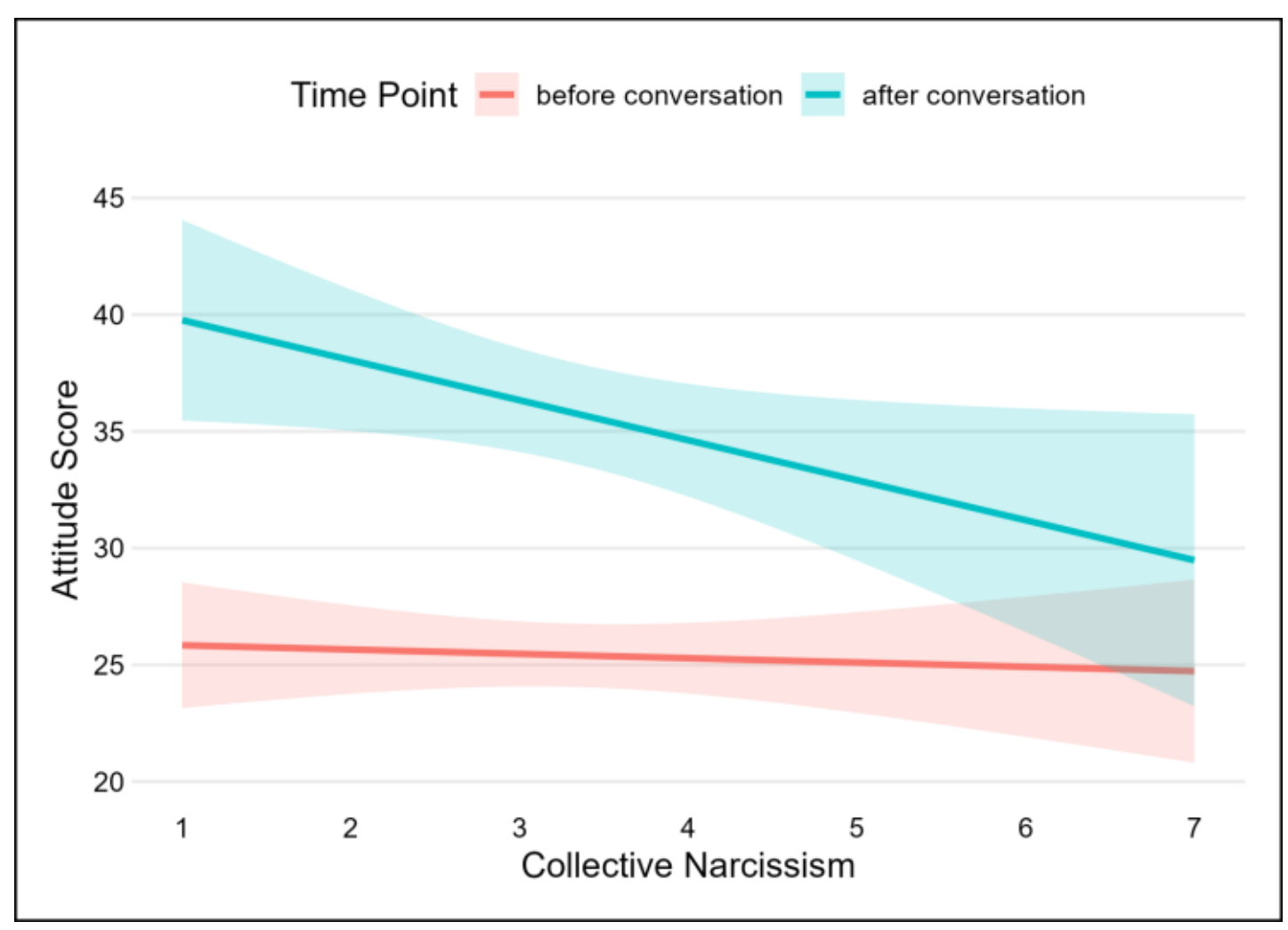

*Note.* The solid lines represent the linear regression fits for attitude scores before (red) and after (blue/green) the conversation across different levels of collective narcissism. The shaded bands surrounding each line indicate the 95% confidence intervals for the predicted means.

**Hypothesis 5: Test of Collective Narcissism × Model Nationality Interaction for Model Perceptions**

We conducted a series of ANCOVAs (general linear model framework) with model nationality (-1, 1) and mean-centered collective narcissism predicting each rating dimension. For all perception variables (perceived objectivity, human-like trust, and functionality trust), the hypothesized nationality × collective narcissism interaction was not significant (see Table 3). This indicates that collective narcissism did not statistically significantly moderate the effect of model nationality on these perceptions.

Given the non-significant interactions across all three perception dimensions, we applied the same regression-based equivalence testing and Bayesian analysis with H4 to confirm the null effects. Results confirmed that the interaction effects were statistically equivalent to zero and strongly supported the null hypothesis for perceived objectivity (TOST $p = .006$, $BF_{01} = 19.09$), human-like trust (TOST $p = .003$, $BF_{01} = 20.04$), and functionality trust (TOST $p = .017$, $BF_{01} = 15.76$).

**Table 3**

*ANCOVA Results for Model Perceptions by Collective Narcissism and Nationality*

| Effect | Perceived objectively | | | | Human-like trust | | | | Functionality trust | | | |
|---|---|---|---|---|---|---|---|---|---|---|---|---|
| | *df* | *F* | *p* | $\eta_p^2$ | *df* | *F* | *p* | $\eta_p^2$ | *df* | *F* | *p* | $\eta_p^2$ |
| Model nationality | 1,399 | 1.97 | .161 | .005 | 1,399 | 6.07 | .014 | .015 | 1,399 | 0.03 | .870 | < .001 |
| Narcissism | 1,399 | 10.83 | .001 | .026 | 1,399 | 44.89 | < .001 | .101 | 1,399 | 11.76 | < .001 | .029 |
| Model nationality × narcissism | 1,399 | 0.10 | .753 | < .001 | 1,399 | 0.00 | .954 | < .001 | 1,399 | 0.48 | .489 | .001 |

## Exploratory Computational Text Analysis

While the pre-registered hypothesis tests relied on pre- and post-conversation self-reports, such static measures treat the persuasive interaction itself as a black box. We therefore analyzed the full transcripts of the 1,209 participant turns (403 participants × 3 rounds) to examine how expressed stance, argumentative conduct, and affective tone evolved over the conversation, and whether these dynamics depended on the model's nationality label and the conversation topic. All analyses in this section are exploratory.

### *Analytic Approach*

**Stance.** Each participant turn was scored on three independent measures of expressed stance. First, as the primary measure, a large language model (GPT-5.5, OpenAI) rated each turn's agreement with the assigned statement (0 = *completely disagrees*, 100 = *completely agrees*) under a constrained JSON output schema; the judge saw only the statement and the single turn, blind to condition, round, and the AI's messages. The same call counted explicit concessions and distinct counter-arguments and flagged whether the turn disparaged the AI or its origin. Second, turns were embedded with a stance-tuned sentence transformer (StanceAware-SBERT; Ghafouri et al., 2024); each sentence was projected onto the axis running from the anti-target to the pro-target statement and projections were averaged within the turn. Third, as a robustness check, the same projection was computed from OpenAI text-embedding-3-large embeddings. All stance scores were aligned so that higher values indicate positions closer to the stance the AI was instructed to advocate; embedding projections were standardized within topic because the two topics' semantic axes differ in discriminability (see Measurement Validity).

**Affect.** Anger, fear, and neutrality were extracted with emotion-english-distilroberta-base (Hartmann, 2022) and toxicity with toxic-bert (Hanu & Unitary team, 2020). Because such classifiers respond to topic vocabulary itself (the non-political statement literally contains the word "stupid") every turn was scored twice: on the raw text and on a masked version in which topic keywords (e.g., "stupid," "surveillance," "national security") were replaced by neutral paraphrases. Comparing raw and masked results separates genuine affect from lexical artifacts.

**Models.** Turn-level scores were analyzed with linear mixed models including round (centered), model nationality (effect-coded: U.S. = −1, China = 1), topic (non-political = −1, political = 1), and all interactions as fixed effects, baseline attitude as a covariate for stance outcomes, and random intercepts and round slopes per participant. Classifier probabilities (i.e., the probability that a classifier assigns to a given emotion or toxicity category for each turn) were logit-transformed and word count was added as a covariate for affect outcomes; *p*-values for the four affect outcomes were Holm-corrected within each effect family. For the absent nationality effects, we additionally computed two one-sided equivalence tests (TOST) and BIC-approximated Bayes factors quantifying evidence for the null.

***Measurement Validity***

The three stance measures converged at the turn level: GPT-coded agreement correlated $r = .55$ with the StanceAware-SBERT projections and $r = .65$ with the OpenAI-embedding projections, and the two embedding measures correlated $r = .73$ (all $n = 1{,}209$, $p < .001$). Validity against the survey was assessed on aligned scales, that is, with both the text measure and the self-report delta coded toward the AI's persuasion goal. Person-level stance displacement (round 3 − round 1) from the GPT judge correlated significantly, if modestly, with self-reported attitude change, whereas the embedding-based displacement did not (Figure 5). We therefore treat the GPT-coded scores as the primary stance measure and the embedding pipelines as convergent robustness checks. Two further caveats motivated this choice. First, correlations computed on direction-preserving (non-aligned) scales are substantially inflated by pooled directional variance and should not be interpreted as validity evidence. Second, using the AI's own turns as known-stance benchmarks showed that the two topics' embedding axes differ sharply in discriminative power (separation between the embedding projections of the AI's turns, which argue the pro-target position, and users' first-round turns, which predominantly oppose it: $d = 1.42$ for the social-media axis vs. $d = 0.37$ for the surveillance axis), so unstandardized embedding scores are not comparable across topics; the GPT rating scale is topic-neutral by construction.

**Figure 5**

*Person-Level Validity of the Text-Based Stance Measures*

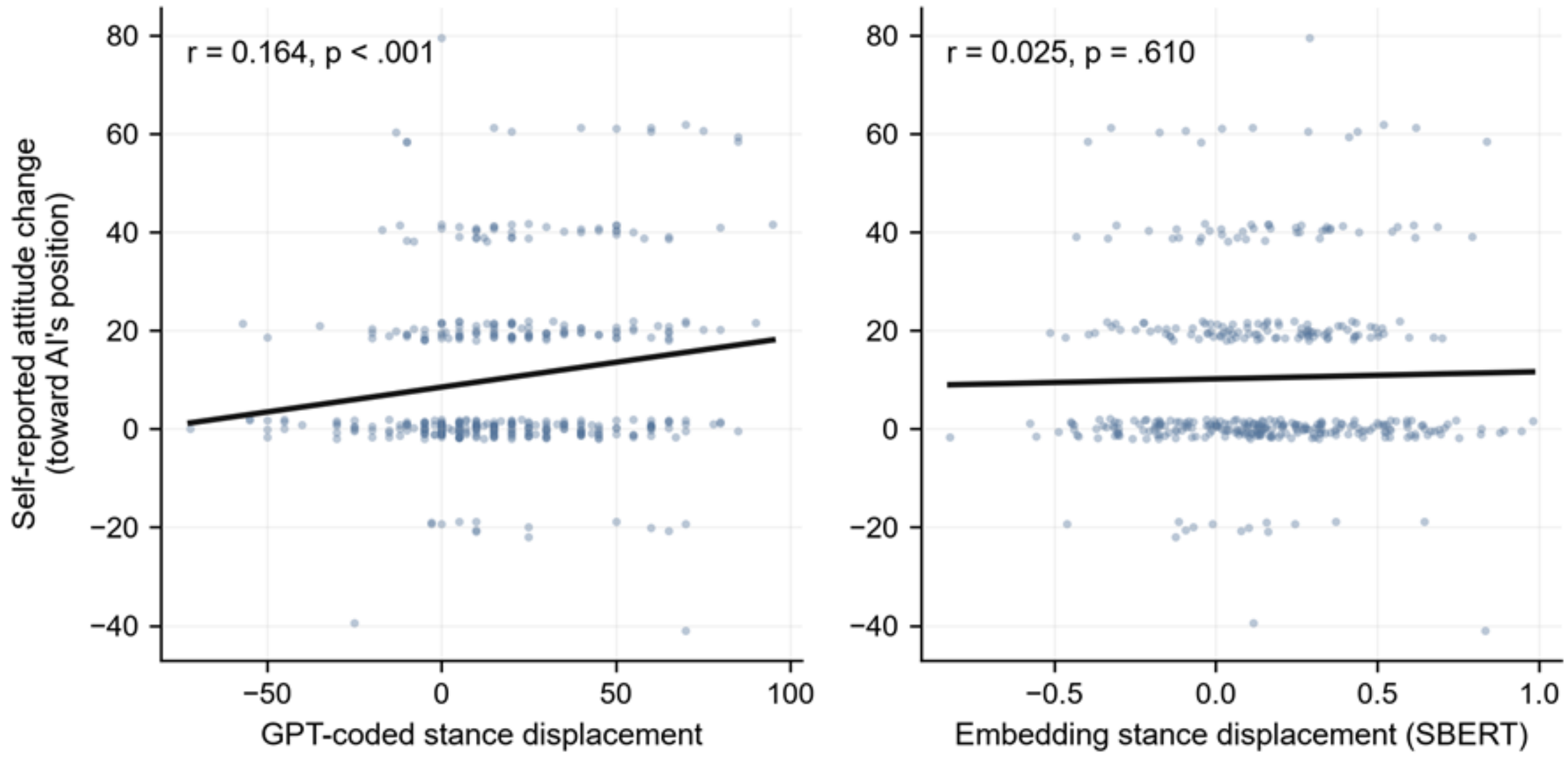

*Note.* Association between self-reported attitude change and stance displacement from the GPT judge (left) and the StanceAware-SBERT embeddings (right), all scales aligned toward the AI's position. Lines are least-squares fits; a small vertical jitter is applied to the discrete survey scores for visibility.

### *Stance Trajectories*

Figure 6 displays the GPT-coded stance trajectories; Figure 7 shows that the identical pattern emerges in both embedding measures, and Figure 8 shows the distribution of individual displacements. Fixed effects for all three growth models are reported in Table 4. Participants' expressed positions moved reliably toward the AI's advocated stance across rounds, corroborating the self-reported persuasion effect (H1) at the behavioral level. Movement was significantly slower in the political than in the non-political conversations, and this moderation replicated in the StanceAware-SBERT and OpenAI-embedding models. Importantly, however, the political trajectories were not flat: expressed stance moved significantly toward the AI's position in every cell of the design.

**Figure 6**

*Expressed Stance Trajectories by Topic and Model Nationality (GPT-Coded)*

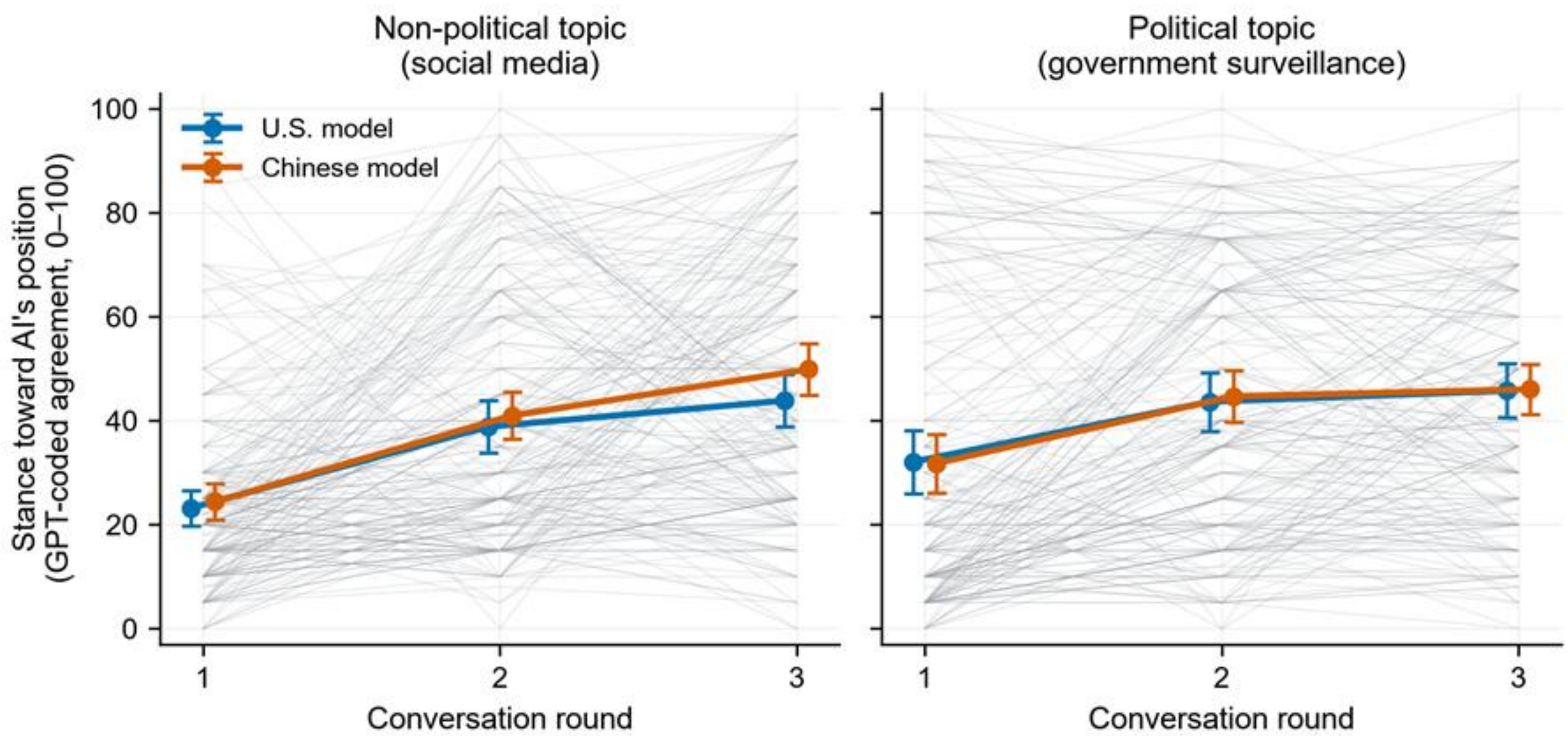

*Note.* Thick lines show condition means with 95% confidence intervals; thin gray lines show all 403 individual participants. Stance is the GPT-5.5-coded agreement of each turn with the assigned statement, aligned so that higher values are closer to the position the AI advocated.

**Figure 7**

*Robustness of the Stance Trajectories Across Three Measurement Pipelines*

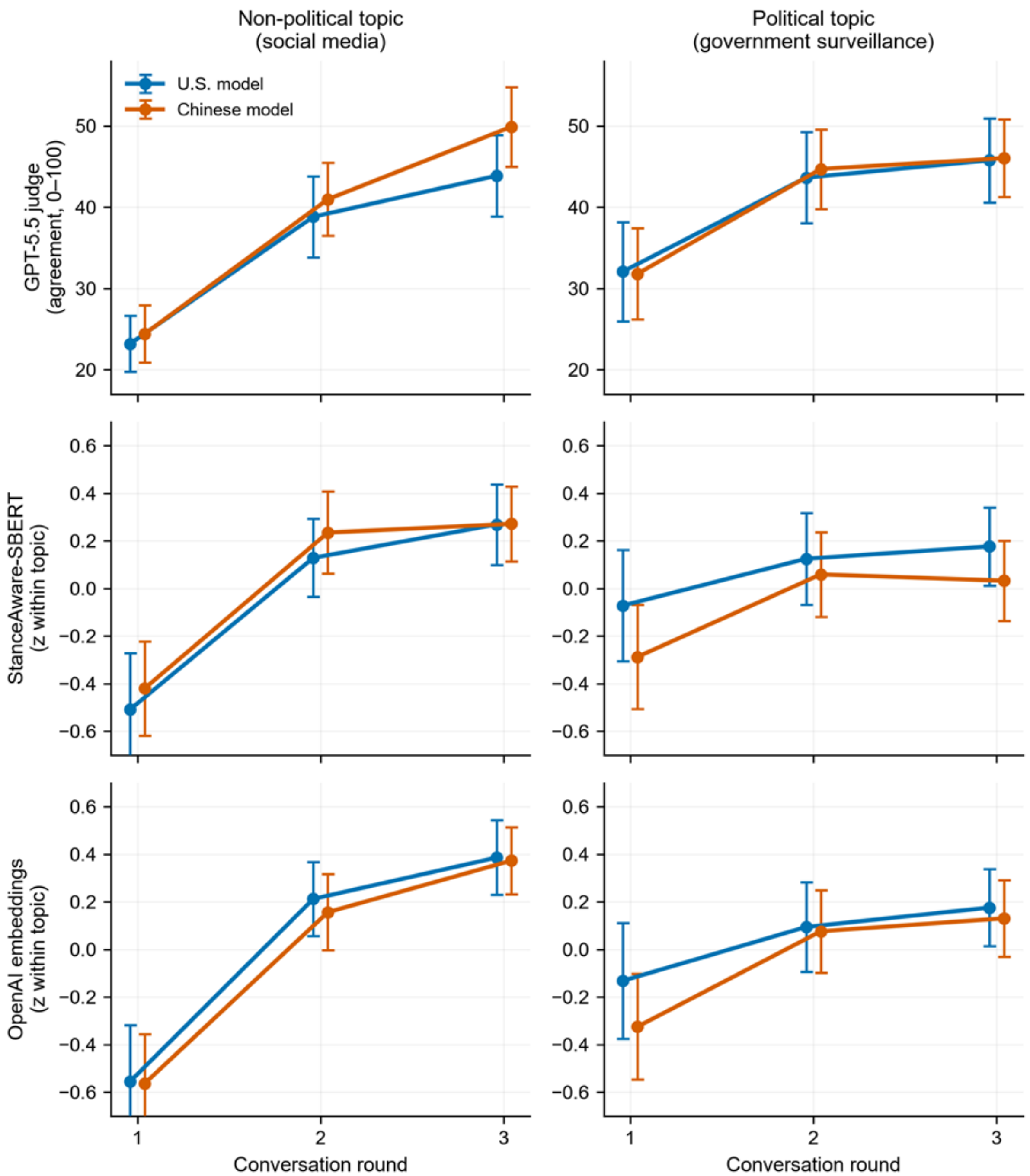

*Note.* Rows show the GPT-5.5 judge ratings, StanceAware-SBERT projections, and OpenAI text-embedding-3-large projections (both embedding measures standardized within topic). Error bars are 95% confidence intervals.

**Figure 8**

*Individual Stance Displacement by Condition*

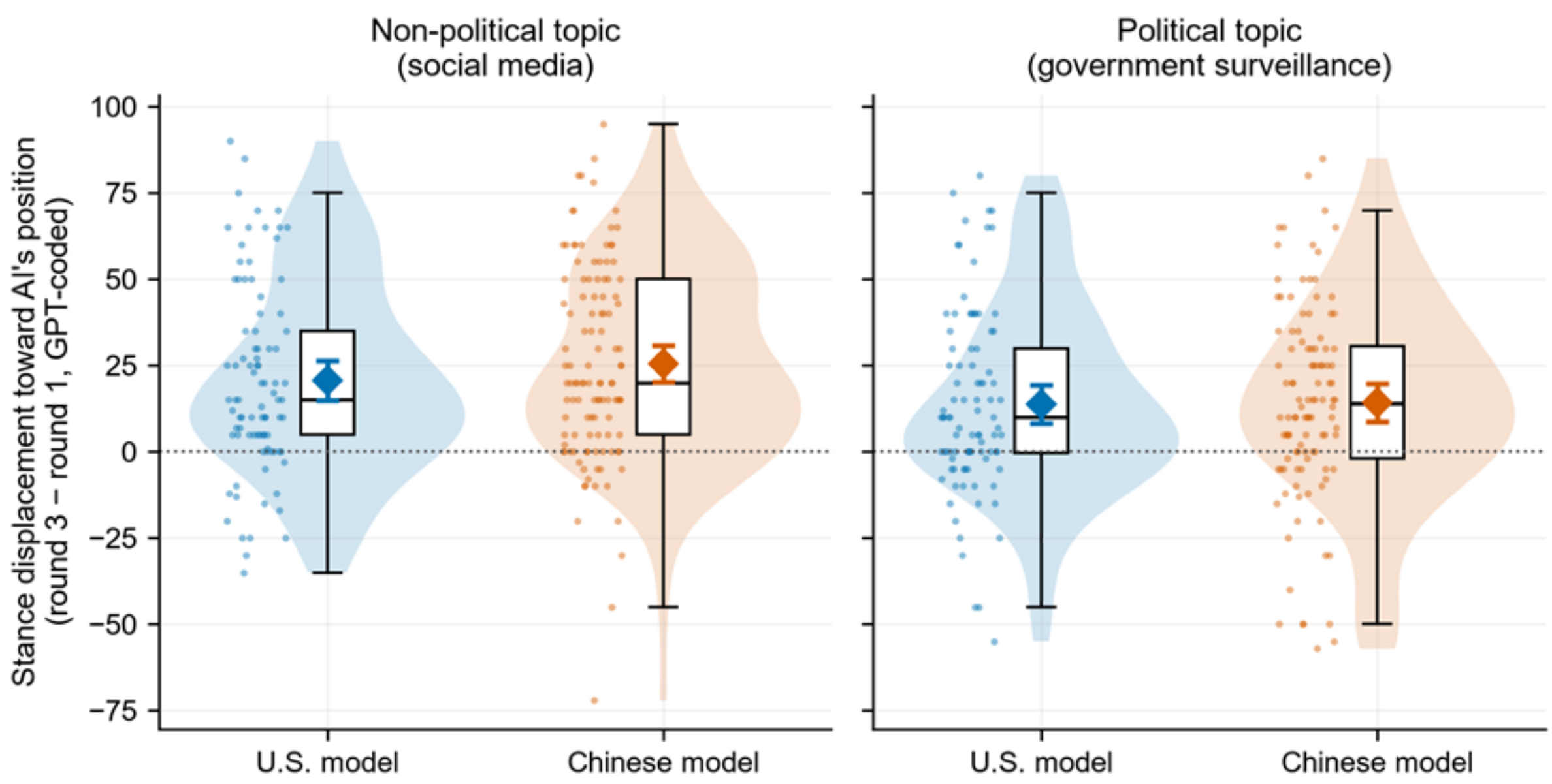

*Note.* Displacement is each participant's GPT-coded stance in round 3 minus round 1. Violins show the distribution; dots show individual participants; white boxes show median and interquartile range; colored diamonds show means with 95% confidence intervals. The dotted line marks zero (no movement).

**Table 4**

*Mixed-Effects Growth Models Predicting Per-Turn Expressed Stance Toward the AI's Position*

| Predictor | *b* [95% CI] | *p* | *b* [95% CI] | *p* | *b* [95% CI] | *p* |
|---|---|---|---|---|---|---|
| | **GPT-5.5 judge** | | **StanceAware-SBERT** | | **OpenAI embeddings** | |
| Intercept | 38.75 [36.86, 40.64] | < .001 | 0.00 [−0.08, 0.08] | .988 | 0.00 [−0.07, 0.08] | .956 |
| Round | 9.26 [7.92, 10.59] | < .001 | 0.25 [0.20, 0.31] | < .001 | 0.33 [0.28, 0.38] | < .001 |
| Nationality (China) | 1.01 [−0.88, 2.91] | .294 | −0.02 [−0.09, 0.06] | .657 | −0.03 [−0.10, 0.05] | .518 |
| Topic (political) | 1.89 [0.00, 3.78] | .050 | 0.01 [−0.07, 0.08] | .902 | 0.00 [−0.07, 0.08] | .986 |
| Round × Nationality | 0.67 [−0.67, 2.01] | .328 | 0.00 [−0.06, 0.05] | .947 | 0.02 [−0.03, 0.07] | .498 |
| Round × Topic | −2.28 [−3.62, −0.94] | .001 | −0.11 [−0.17, −0.06] | < .001 | −0.14 [−0.19, −0.09] | < .001 |
| Nationality × Topic | −0.86 [−2.75, 1.04] | .374 | −0.05 [−0.13, 0.02] | .158 | −0.02 [−0.09, 0.06] | .634 |
| Round × Nationality × Topic | −0.53 [−1.87, 0.81] | .436 | 0.02 [−0.03, 0.07] | .465 | 0.02 [−0.03, 0.07] | .470 |
| Baseline attitude | 4.84 [2.96, 6.73] | < .001 | 0.07 [−0.01, 0.14] | .068 | 0.10 [0.03, 0.17] | .006 |

*Note.* $N$ = 403 participants, 1,209 turns. All models include random intercepts and random round slopes per participant. Nationality and topic are effect-coded (−1, 1); round is centered; baseline attitude is standardized. The GPT-judge outcome is the 0–100 agreement rating aligned toward the AI's position; embedding outcomes are axis projections standardized within topic. $p$ values are Wald tests.

In contrast, the nationality label had no detectable influence on the stance dynamics. Neither the main effect of nationality, nor the round × nationality interaction, nor, critically for the theoretical framework, the round × nationality × topic interaction approached significance in any of the three measures (see Table 4). Equivalence testing indicated that this is an informative null rather than mere absence of evidence (see Table 5): in the full sample, the China–U.S. difference in text-based stance displacement was statistically equivalent to zero even within narrow bounds, and the Bayes factors provided strong evidence for the null overall and within each topic.

**Table 5**

*Equivalence Tests and Bayes Factors for China–U.S. Differences in Conversational Behavior*

| Outcome | Sample | $d$ | TOST $p$ | $BF_{01}$ |
|---|---|---|---|---|
| Stance displacement | Full sample | −0.00 | .003 | 20.1 |
| | Non-political topic | −0.07 | .070 | 12.5 |
| | Political topic | 0.08 | .077 | 12.2 |
| Concessions per turn | Full sample | −0.04 | .010 | 18.4 |
| Counter-arguments per turn | Full sample | −0.23 | .304 | 1.5 |
| Toxicity (masked) | Full sample | 0.03 | .005 | 19.1 |
| Anger (masked) | Full sample | 0.19 | .189 | 3.1 |
| Fear (masked) | Full sample | −0.11 | .042 | 11.5 |
| Neutrality (masked) | Full sample | 0.00 | .003 | 20.1 |

*Note.* Stance displacement is the GPT-coded stance in round 3 minus round 1; the remaining outcomes are participant means across the three turns, with affect scores taken from the keyword-masked text. Positive $d$ indicates a higher value in the China condition. TOST bounds are $d = \pm 0.28$, the smallest effect the study was powered to detect (the full-sample displacement result also holds at $d = \pm 0.20$, TOST $p = .024$); $BF_{01}$ is the BIC-approximated Bayes factor in favor of no difference (values > 3 moderate, > 10 strong evidence for the null).

***Argumentative Conduct***

The GPT-coded conduct measures (see Figure 9) showed that explicit concessions roughly tripled from the first to the later rounds ($b$ = 0.20 concessions per turn per round, 95% CI [0.16, 0.24], $p < .001$), with no difference by nationality ($b$ = −0.01 [−0.05, 0.03], $p$ = .648), for which equivalence was supported (Table 5), and only a marginal tendency toward fewer concessions in political conversations ($b$ = −0.04 [−0.08, 0.00], $p$ = .058). Counter-arguing showed small, uncorrected effects of nationality ($b$ = −0.09 [−0.16, −0.01], $p$ = .019), with fewer counter-arguments against the Chinese model, opposite to the resistance hypothesis, and topic ($b$ = −0.08 [−0.16, −0.01], $p$ = .026), and increased over rounds in political relative to non-political conversations ($b$ = 0.08 [0.02, 0.13], $p$ = .007). None of these effects survives Holm correction, and for counter-arguing neither a difference nor equivalence could be established ($BF_{01}$ = 1.5; Table 5), so these trends should not be interpreted substantively.

**Figure 9**

*GPT-Coded Concessions and Counter-Arguments per Turn by Condition*

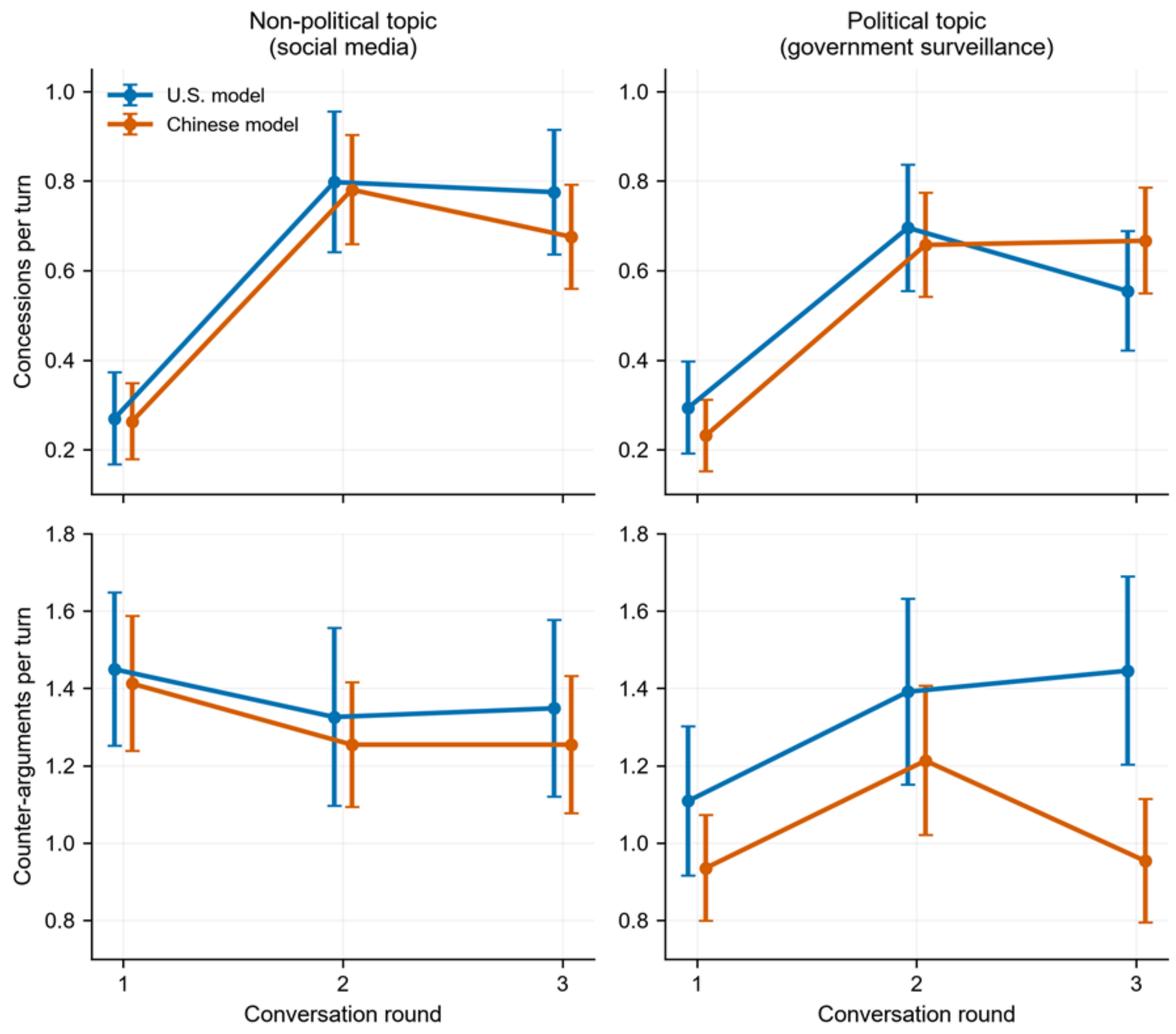

*Note.* Means with 95% confidence intervals across conversation rounds, by topic (panels) and model nationality (colors).

Most strikingly, overt hostility toward the agent was nearly absent. Across all 1,209 argumentative turns (roughly half of them addressed to an AI explicitly and repeatedly labeled as the product of a geopolitical rival) only 10 turns (0.8%) contained any disparagement of the AI or its origin, and these were distributed proportionally across the U.S. (4/543, 0.7%) and Chinese (6/666, 0.9%) conditions. Whatever reservations participants held about the out-group model, recall the lower human-like trust reported before the conversation, they did not surface as source derogation during the interaction itself.

**Affective Dynamics**

The keyword-masking comparison (see Figure 10; Table 6) revealed that the toxicity patterns in the raw text were largely lexical artifacts. In raw text, toxicity was far higher in the non-political topic ($b = -0.64$ on the logit scale, 95% CI [−0.77, −0.51], $p < .001$) and declined steeply over rounds ($b = -0.52$ [−0.66, −0.39], $p < .001$); with topic keywords masked, the topic effect shrank by roughly 90% ($b = -0.08$ [−0.15, −0.01], $p = .033$) and the decline disappeared entirely ($b = 0.01$ [−0.07, 0.08], $p = .879$). Both patterns thus reflect participants restating and then ceasing to restate the assigned phrase “makes people stupid,” not genuine hostility. A focused test of the apparent first-round toxicity difference between the Chinese and U.S. models in the non-political topic was non-significant once word count was controlled (raw: $b = 0.023$ [−0.011, 0.057], $p = .190$; masked: $b = 0.001$ [−0.002, 0.004], $p = .412$).

**Figure 10**

*Toxicity and Anger Trajectories in Raw Versus Keyword-Masked Text*

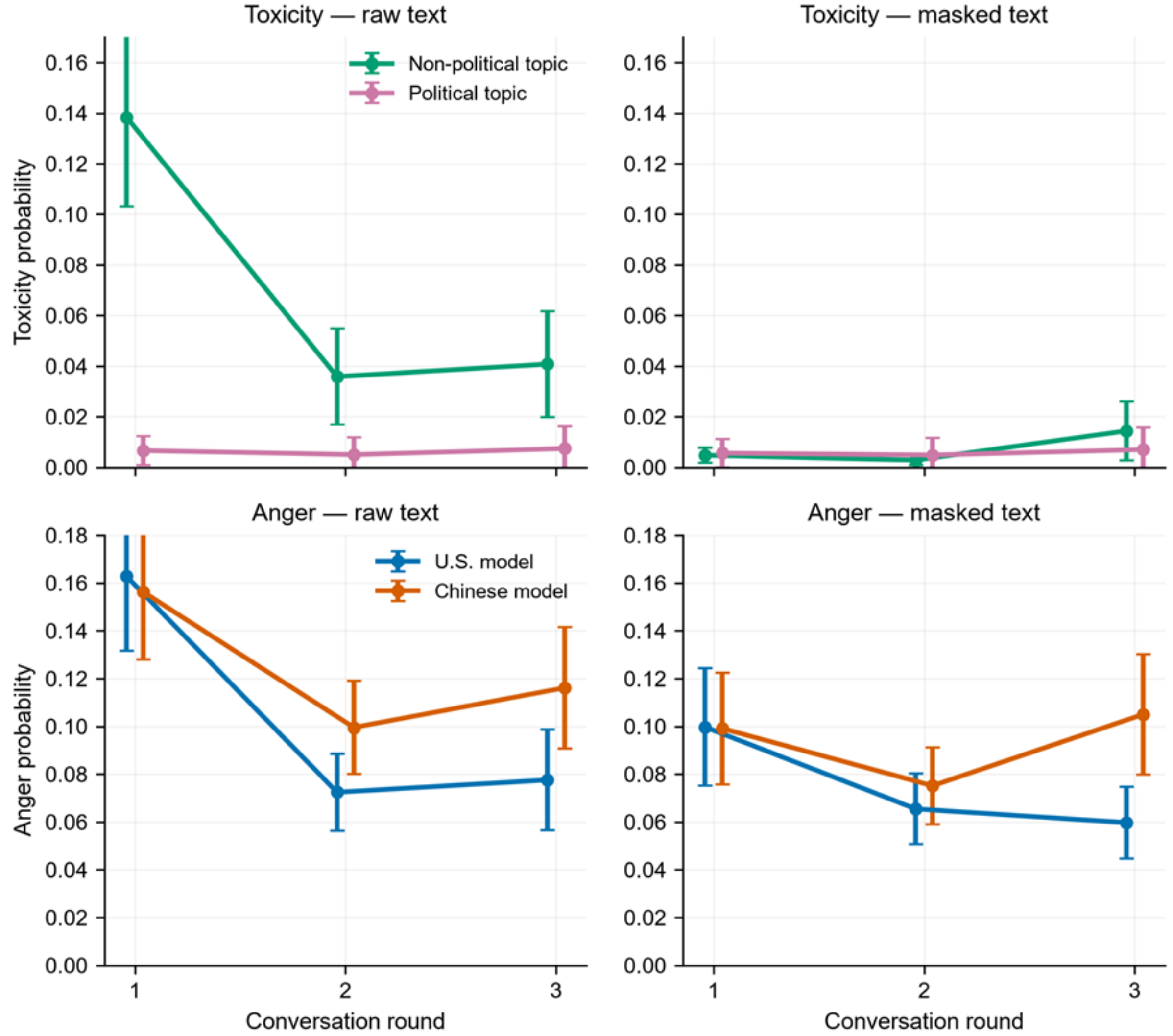

*Note.* Means with 95% confidence intervals. Masking replaces topic keywords (e.g., “stupid,” “surveillance”) with neutral paraphrases; the near-elimination of the toxicity effects under masking identifies them as lexical artifacts. Y-axis scales are identical within each row.

**Table 6**

*Fixed Effects of Topic, Round, and Nationality on Affective Outcomes in Raw Versus Keyword-Masked Text (Logit Scale)*

| Term | Outcome | Raw $b$ | Raw $p_{\text{Holm}}$ | Masked $b$ | Masked $p_{\text{Holm}}$ |
|---|---|---|---|---|---|
| Topic (political) | Toxicity | −0.64 | < .001 | −0.08 | .033 |
| | Anger | 0.02 | .672 | 0.30 | < .001 |
| | Fear | 1.10 | < .001 | 0.95 | < .001 |
| | Neutrality | −0.32 | < .001 | −0.48 | < .001 |
| Round | Toxicity | −0.52 | < .001 | 0.01 | .879 |
| | Anger | −0.36 | < .001 | −0.12 | .057 |
| | Fear | −0.27 | < .001 | −0.11 | .198 |
| | Neutrality | 0.48 | < .001 | 0.20 | .017 |
| Round × Topic | Toxicity | 0.41 | < .001 | −0.05 | .333 |
| | Anger | 0.12 | .062 | −0.07 | .333 |
| | Fear | −0.35 | < .001 | −0.27 | < .001 |
| | Neutrality | 0.00 | .962 | 0.12 | .230 |
| Nationality (China) | Toxicity | 0.07 | .996 | −0.01 | 1.00 |
| | Anger | 0.10 | .189 | 0.10 | .133 |
| | Fear | −0.02 | 1.00 | 0.00 | 1.00 |
| | Neutrality | 0.02 | 1.00 | 0.00 | 1.00 |
| Round × Nationality | Toxicity | −0.10 | .403 | −0.06 | .358 |
| | Anger | 0.10 | .303 | 0.09 | .291 |
| | Fear | 0.06 | .806 | 0.00 | 1.00 |
| | Neutrality | −0.06 | .806 | −0.03 | 1.00 |
| Round × Nationality × Topic | Toxicity | 0.08 | 1.00 | 0.02 | 1.00 |
| | Anger | 0.01 | 1.00 | 0.03 | 1.00 |
| | Fear | 0.05 | 1.00 | 0.01 | 1.00 |

| | | | | |
|---|---|---|---|---|
| Neutrality | −0.04 | 1.00 | −0.01 | 1.00 |

*Note.* Linear mixed models on logit-transformed classifier probabilities with word count as covariate and random intercepts and round slopes per participant. $p$ values are Holm-corrected across the four outcomes within each term and text version. Masked text replaces topic keywords (e.g., "stupid," "surveillance") with neutral paraphrases.

Anger showed a weak, non-robust nationality pattern. Anger declined over rounds in conversations with the U.S. model but not with the Chinese model; however, the critical round × nationality interaction was not significant in either text version (Table 6), the small overall elevation of anger toward the Chinese model did not survive Holm correction, and equivalence testing for anger was inconclusive (Table 5). These trends are therefore reported for transparency but do not license the conclusion of sustained emotional resistance toward the out-group model. By contrast, two affective signatures were robust to masking: fear was substantially higher in political conversations and decreased over rounds primarily there, and linguistic neutrality rose steadily over rounds in all conditions (Table 6). For masked toxicity, fear, and neutrality, the China–U.S. differences were statistically equivalent to zero (Table 5). No nationality term reached corrected significance for any affect outcome, including all round × nationality × topic interactions (all Holm-corrected $p = 1.0$).

## Discussion

LLMs demonstrate a substantial capacity to persuade and alter human attitudes during interactive debates, aligning with findings from our nationally representative sample. As predicted, individuals reported significant attitude shifts following multi-turn conversations with the artificial agents. Contrary to our predictions, however, the explicit national origin label of the model did not significantly moderate the magnitude of the final self-reported attitude change. Importantly, our computational analysis of the conversation transcripts converged with this conclusion. Participants' expressed stance moved toward the AI's position in every condition, and neither the trajectory of this movement, nor participants' argumentative conduct, nor their affective tone differed by the model's alleged nationality, with equivalence tests and Bayes factors indicating that these are informative null effects. The absence of an out-group penalty in the final attitude scores thus does not appear to conceal hidden behavioral resistance. Rather, resistance specific to the out-group model failed to materialize at any level we measured, with one notable exception: participants extended less human-like trust to the Chinese model before the conversation even began.

To understand why the national label might not have altered the ultimate attitude scores, it is helpful to examine the specific cognitive pathways through which users evaluate the AI. The differential correlates we observed validate that respondents actively distinguished between distinct psychological dimensions of trust (Choung et al., 2023). Specifically, the out-group national label significantly diminished human-like trust but showed no statistical impact on functionality trust. Because users may fundamentally approach AI as a utilitarian tool, this preserved functionality trust could have functioned as the primary driver of the persuasive outcome, potentially compensating for the user's skepticism regarding the system's benevolence. Furthermore, post-conversation assessments of perceived objectivity revealed no significant differences across the national conditions. This pattern suggests that both the a priori impression of the system's capability and the actual conversational experience might have led users to perceive the tool as functionally usable and logically sound, irrespective of its geopolitical origin. Consequently, the national label did not systematically disrupt the explicit post-test agreement scores. We speculate that these biases were not detected by traditional statistical tests of final attitudes because users were able to separate the tool's practical utility from its national origin. This interpretation is reinforced by the transcripts themselves. Participants engaged with the two models in a virtually indistinguishable manner, conceding, counter-arguing, and modulating their emotions at comparable rates, which suggests that the withheld social trust remained a declarative reservation that never translated into behavioral resistance during the interaction.

Our third hypothesis, predicting that perceived objectivity and the two dimensions of trust in AI would mediate the relationship between model nationality and attitude change, was not supported. However, our subsequent path analysis revealed a nuanced dynamic. We found that trust in AI failed to predict post-conversation attitudes in three of the four experimental conditions. By contrast, within the Chinese-political condition, the two trust dimensions uniquely predicted the attitude shift, and they did so in opposite directions. Functionality trust positively predicted movement toward the AI's position, whereas human-like trust negatively predicted it. The positive path is the less surprising of the two. Functionality trust captures an appraisal of the system as a tool, and participants who judged the tool competent may simply have given its arguments a fairer hearing, evaluating them on their merits rather than dismissing them at the source.

The negative path requires more interpretation. We propose that the two trust dimensions correspond to two different construals of the same agent. Assessing a model's functionality means treating it as an instrument, whereas assessing its benevolence and integrity means treating it as a social actor with intentions of its own. For a domestic model, or in a benign discussion, this anthropomorphized construal is inconsequential. When the agent carries an out-group label and the topic is political, however, granting its intentions invites the question of whose interests those intentions serve. A foreign AI perceived as "sincere" and "principled" may, by implication, be sincerely and faithfully committed to the values of its creators. On this reading, high human-like trust did not indicate comfort with the model; it indicated that the participant had construed the model as an agent of a rival country, which is precisely the construal under which social identity theory predicts resistance to persuasion. Participants who kept the foreign model in the tool category were moved by its arguments, while those who elevated it to the status of a social agent defended their position against it. This would also explain why the pattern appeared only in the Chinese-political condition. It is the combination of an out-group source and an identity-relevant topic that turns anthropomorphization from a neutral attribution into a threat cue. We stress that this pattern emerged from an exploratory, condition-specific analysis of correlational paths estimated after the hypothesized mediation model was not supported. It should therefore be read as a hypothesis for future research on when anthropomorphizing foreign AI systems helps or hinders persuasion, not as an established mechanism.

Turning to individual differences, we did not find evidence for our hypothesis that collective narcissism would specifically moderate the effect of model nationality. Instead of uniquely resisting the out-group AI, collective narcissism appeared to inhibit persuasion across all experimental conditions. Participants scoring higher in collective narcissism exhibited significantly less attitude change overall. This suggests that high collective narcissists might not merely be sensitive to out-group threats, but rather, they could possess a high degree of defensiveness against any external force attempting to alter their established beliefs, regardless of whether that force is a domestic or foreign AI. We speculate that this generalized resistance could be explained by the underlying link between individual and collective narcissism (Golec de Zavala et al., 2013). Because individual narcissism is often correlated with collective narcissism, and narcissistic individuals are generally less open to changing their opinions, this rigid adherence to existing viewpoints might translate into a universal shield against algorithmic persuasion. By illustrating that this trait functions as a general barrier to persuasion rather than solely an out-group filter, these findings potentially expand current theoretical understandings of collective narcissism in digital environments.

**Unpacking the Persuasion Black Box**

Diverging from the self-report analyses in granularity but not in substance, the computational analysis corroborated the persuasion effect at the behavioral level and clarified where resistance did and did not occur. Participants' expressed positions moved reliably toward the AI's advocated stance across rounds in all three measurement pipelines, and explicit concessions roughly tripled from the first to the later rounds. Persuasion was thus visible not only in the pre-post survey deltas but in the moment-to-moment language of the debate itself.

The conversational topic, not the model's nationality, shaped these dynamics. Expressed stance moved more slowly toward the AI's position in political than in non-political conversations, a moderation that replicated across all three stance measures, and political exchanges were marked by substantially higher levels of fear that subsided over rounds. Notably, the political trajectories were not flat. The expressed stance moved significantly toward the AI's position in every cell of the design. Rather than blanket resistance, this pattern suggests a more guarded style of engagement in identity-relevant domains, in which users yield ground verbally at a slower rate even as they update their positions.

By contrast, the nationality label left no detectable trace in the conversations. Stance trajectories, concessions, counter-arguing, and affective tone were statistically equivalent across the U.S. and Chinese conditions, and overt disparagement of the agent was nearly absent, occurring in under one percent of turns and distributed similarly across conditions. An apparent elevation of toxicity toward the Chinese model, along with a seemingly persistent anger trajectory, did not survive keyword masking, word-count control, or correction for multiple comparisons. Indeed, the masking procedure revealed that the most striking affective patterns in the raw text were lexical artifacts of participants restating the assigned topic phrases.

Taken together, the behavioral record indicates a clear dissociation: participants reported less human-like trust in the out-group model when asked directly, yet nothing in their actual conversational conduct distinguished their treatment of the two agents. Within the CASA framework, users evidently registered the social identity cue, but the cue did not activate the defensive repertoire, such as counter-arguing, source derogation, or emotional escalation, that intergroup theory would predict in human-to-human persuasion.

**Limitations and Future Directions**

Unlike recent findings suggesting that perceived partisan bias in LLMs significantly reduces their persuasive abilities (DiGiuseppe & Robison, 2026), we did not observe a significant impact of the national label on explicit attitude change. We speculate that this divergence might occur because, within a representative United States sample, the geopolitical "Chinese" label might not universally evoke the same visceral, immediate hostility as domestic partisan polarization. Geopolitical competitors, while strategically significant, could be perceived by the general public as more abstract or distant than the highly polarized, immediate domestic political climate.

Several further limitations qualify our conclusions. First, the computational findings rest on exploratory analyses conducted with off-the-shelf classifiers whose error characteristics on conversational, first-person text are imperfectly known, and the keyword-masking results themselves illustrate how sensitive such tools are to surface features of the input. Similarly, although the LLM-based stance coding was validated against self-reported attitude change and converged with two independent embedding pipelines, it remains a model-based measure whose judgments cannot simply be equated with human annotation.

Second, the interaction was brief. Three conversational rounds may suffice to detect immediate persuasion but may be too short for slower-building forms of resistance, such as accumulating reactance or fatigue, to surface in participants' language.

Third, our manipulation varied only the label attached to an otherwise identical model. Because both conditions relied on GPT-4o, all participants in fact encountered the rhetorical style of a model developed in the United States that they may be familiar with, and the conversations carried none of the cultural intricacies, argumentation styles, or value emphases that an actually Chinese-developed model might display. Our design therefore cleanly isolates the effect of perceived origin, but it cannot speak to how users respond to the authentic linguistic and cultural signatures of foreign systems. Relatedly, the model behaved uniformly politely and cooperatively across conditions. Users' indifference to the origin label may therefore be bounded to such benign conduct; if an AI were to argue manipulatively, apply pressure, or assert false claims, its perceived national origin might become a salient cue that reactivates intergroup vigilance. Future research should test whether origin effects emerge once the agent's behavior itself arouses suspicion.

Fourth, although controversial, the topics were not geopolitically implicated in the relationship between the two countries. A debate about domestic surveillance or social media does not touch on issues where American and Chinese interests visibly collide, such as trade, Taiwan, or technological competition. An out-group label may carry considerably more weight when the persuader argues a position from which its country of origin stands to benefit, and future research should test whether the null effects observed here extend to such self-serving persuasion contexts.

Finally, due to financial and temporal constraints, we were unable to conduct a reciprocal experiment within China, nor did we introduce a third-party benchmark, such as an AI model from Europe. Future research should strive to cross-culturally validate these findings by examining how different populations react to allied versus adversarial foreign algorithms.

## Conclusion

In an era where AI increasingly serves as a primary conduit for information and public discourse, understanding the psychological boundaries of algorithmic persuasion is paramount for safeguarding democratic integrity and information sovereignty. Our research replicates that the persuasive power of LLMs is substantial and, more strikingly, extends this finding by showing that it is largely indifferent to the machine's perceived geopolitical identity. Participants updated their attitudes, conceded points, and softened their language at comparable rates whether they believed they were debating an American or a Chinese system. The national label left a single reliable trace, a reluctance to grant the foreign model human-like trust, and even this reservation did not translate into behavioral resistance or diminished persuasion. These findings carry sobering implications for global technology governance. If the mere disclosure of a rival origin neither blunts an AI's arguments nor provokes detectable defensiveness during the interaction, then origin labeling and transparency requirements alone may offer weak protection against foreign influence operations conducted through conversational AI. While AI can effortlessly cross physical borders, the psychological borders that citizens were expected to defend appear far more permeable than intergroup theory would predict. Anticipating the societal impact of sovereign AI will therefore require looking beyond users' spontaneous defenses toward structural safeguards for transparent and constructive human-AI interaction on a global scale.

## Author Statement

The questionnaires, code, and data associated with this study are available on OSF (https://osf.io/ts7g6/overview?view_only=8c56bbba10ea4b55ab13c3cdb3ba9688). LLMs were used to assist in code debugging and manuscript proofreading, specifically focusing on detecting spelling, grammatical errors, and awkward English phrasing. We assume full responsibility for the final content.

## Funding

This work was supported by funding provided to Ningzhi Liu from University of Oslo's UiO:Democracy program.

## Data Availability

The data supporting the findings of this study are openly available on the Open Science Framework at https://osf.io/8ba2c/overview?view_only=8d690e2a74c94bbd9cea3d9a3336777e

## Author Contribution

Ningzhi Liu: Conceptualization, Methodology, Investigation, Data curation, Formal analysis, Visualization, Writing – original draft. Yannic Hinrichs: Conceptualization, Methodology, Formal analysis, Writing – review & editing. Jonas R. Kunst: Conceptualization, Methodology, Formal analysis, Validation, Visualization, Writing – original draft, Writing – review & editing, Supervision. All authors read and approved the final version of the manuscript.